\documentclass[aps,prl,twocolumn,superscriptaddress]{revtex4}

\usepackage{natbib}
\usepackage{graphicx}
\usepackage{dcolumn}
\usepackage{bm}

\begin{document}

\title{Spin excitations of the correlated semiconductor FeSi
probed by THz radiation}

\author{V. V. Glushkov}
 \email{glushkov@lt.gpi.ru}
 \altaffiliation{Also at: Moscow Institute of Physics and Technology, 9 Institutskii per., Dolgoprudnyi, Moscow Region 141700 Russia}
\author{B. P.~Gorshunov}
 \altaffiliation{Also at: Moscow Institute of Physics and Technology, 9 Institutskii per., Dolgoprudnyi, Moscow Region 141700 Russia}
\author{E.~S.~Zhukova}
 \altaffiliation{Also at: Moscow Institute of Physics and Technology, 9 Institutskii per., Dolgoprudnyi, Moscow Region 141700 Russia}
\author{S.~V.~Demishev}
 \altaffiliation{Also at: Moscow Institute of Physics and Technology, 9 Institutskii per., Dolgoprudnyi, Moscow Region 141700 Russia}
\author{A. A. Pronin}
\author{N. E. Sluchanko}
 \affiliation{A. M. Prokhorov General Physics Institute of RAS, 38,
Vavilov str., Moscow, 119991, Russia}
\author{S. Kaiser}
\author{M. Dressel}
 \affiliation{1. Physikalisches Institut, Universit\"at Stuttgart, Pfaffenwaldring 57, 70550 Stuttgart, Germany}

\date{\today}

\begin{abstract}
By direct measurements of the complex optical conductivity $\sigma(\nu)$ of FeSi we have discovered a broad absorption peak centered at frequency $\nu_{0}(4.2~{\rm K}) \approx 32~{\rm cm}^{-1}$ that develops at temperatures below $20$~K. This feature is caused by spin-polaronic states formed in the middle of the gap in the electronic density of states.  
We observe the spin excitations
between the electronic levels split by the exchange field of
$H_{e}=34\pm 6$~T. Spin fluctuations are identified as the main factor 
determining the formation of the spin polarons
and the rich magnetic phase diagram of FeSi.
\end{abstract}


\maketitle

The origin of the narrow gap ($E_\mathrm{g}\sim60$~meV) that develops in the density of electronic states of FeSi at low temperatures poses a real challenge to both theorists and experimentalists for last five decades \cite{Aeppli92,Riseborough00}. Due to the anomalous transfer of spectral weight from the gap region to high energies (up to $10 E_\mathrm{g}$) \cite{Schles93,Degiorgi94,Mena06} FeSi was put in line with the Kondo insulators Ce$_{3}$Bi$_{4}$Pt$_{3}$, SmB$_{6}$, YbB$_{12}$ {\it etc.} At the same time, the crossover from an activated to a Curie-Weiss behavior of the magnetic susceptibility \cite{Jacc67} was successfully explained within a spin-fluctuation model suggesting thermally induced magnetic moments of Fe \cite{Takahashi97}. Finally, the distinct $d$-symmetry of electronic states above and below the gap \cite{Fu95} makes it important to consider carefully the effects of strong electron correlations \cite{Anisimov96,Maz10,Franco07,Franco09}.

The interest to FeSi has been renewed after recent high resolution photoemission experiments, which show no sign of a Kondo resonance favoring  a simple itinerant semiconductor picture with $d-$bands renormalized due to correlation effects \cite{Zur07,Klein08}. 
Ellipsometric studies in a broad spectral range supports this assumption \cite{Menzel2009}; the collapse of the gap is caused by a temperature-induced transition from a low-temperature semiconductor to a high-temperature semimetal. However, this conclusion is at variance with the results of dc-transport and magnetic studies \cite{Sluchanko2000,Sluchanko2002,Corti03} which assign the onset of the metallic state below 80~K to the formation of a narrow many-body resonance (width $E_p\sim 6$~meV) within the gap. The tiny but finite residual intensity at the Fermi level was also found in low-energy photoemission spectra \cite{Arita2008}. Hence, it is particularly important to identify the intra-gap features that are responsible for the anomalous ground state properties of FeSi and to compare them with the prediction of existing theories \cite{Anisimov96,Maz10,Franco07,Franco09,Jarlborg07}.

In this Letter we report on the first direct measurements of the complex optical conductivity $\sigma(\nu)=\sigma_{1}+{\rm i}\sigma_{2}$ of FeSi in the low-frequency range $8-40~{\rm cm}^{-1}$, supplemented by broad-range experiments in an unprecedented large spectral range from 10 Hz to $1.2\times 10^{15}$~Hz.  
We identify a spin-polaronic resonance in the middle of the energy gap that is split by the exchange field $H_{e}=34\pm 6$~T.  Excitations  
between these electronic levels lead to an anomalous contribution to $\sigma(\nu)$ around  $\nu_{0}=32~{\rm cm}^{-1}$. From the comparison between the intra-gap excitation parameters of FeSi, SmB$_{6}$ and YbB$_{12}$ \cite{Gorshunov99,SluchPRB00,Gorshunov06} we conclude that the low-temperature optical properties of these compounds cannot be accounted for within the Kondo insulator model.

Single crystals of FeSi under investigation were grown by Czochralski technique. The samples were cut and subsequently polished by abrasive diamond powder. The fragility of the crystals constituted a limit of how thin the samples could be polished (about 35 to 50~$\mu$m). In order to remove the surface defects induced by polishing, the specimens were etched in the mixture of hydrofluoric, nitric and acetic acids taken at the ratio of 2:3:2. Note, $\sigma_\mathrm{dc}$ of polished crystals exceeds that of etched samples by more than two orders of magnitude (Fig.~\ref{Fig1}), we limited our optical studies to thoroughly etched samples.

A quasioptical THz spectrometer based on backward wave oscillators \cite{Gorshunov99} was employed to measure the transmission and phase spectra in the spectral range from 8 to 40~cm$^{-1}$ at temperatures $4~{\rm K}<T<20$~K. From these quantities the complex optical conductivity is calculated. Additional  infrared reflection measurements were conducted with a Bruker IFS 113V ($50-10^{4}~{\rm cm}^{-1}$) complemented by a spectrometric ellipsometer (Woollam VASE, up to $4\times 10^{4}~{\rm cm}^{-1}$). The microwave conductivity was measured by cavity-perturbation technique at fixed frequencies 9.5, 35 and 100~GHz using Gunn diodes as radiation sources. Radiofrequency investigations ($10-1000$~MHz) were carried out utilizing  a HP Impedance Analyzer HP4191A. Standard four-probe technique was applied to obtain $\sigma_\mathrm{dc}(T)$ used as reference data at zero frequency. The Kramers-Kronig analysis of the infrared reflectivity spectra was performed by utilizing the directly measured values of $\sigma_{1}$ and $\sigma_{2}$ below and above the infrared range. Eventually we obtained a very wide panorama of the optical conductivity $\sigma_{1}$($\nu$) and of the dielectric function $\epsilon_1(\nu)=1-4\pi\sigma_{2}(\nu)/\nu$.

\begin{figure}
\includegraphics[width=1\linewidth]{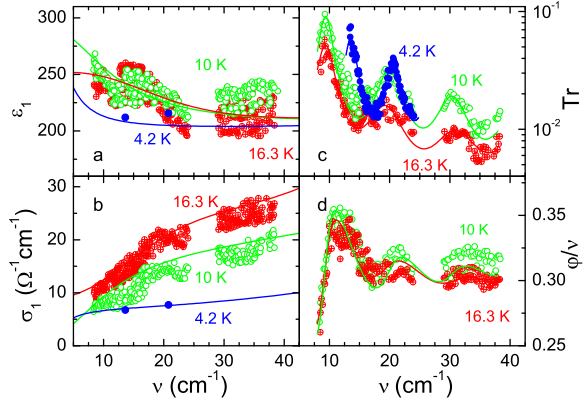}%
\caption{\label{Fig2} (Color online) Terahertz spectra of (a) optical conductivity $\sigma_1$ and (b) dielectric function $\epsilon_1$ of FeSi calculated from (c) transmission $Tr$ and (d) normalized phase shift $\varphi/\nu$ at $T=4.2$~K, 10~K and 16.3~K (blue, green, and red symbols, respectively). Solid lines present the results of the least-square fitting by Eq.\ref{Eq1} (a,b) or by interferometric equations (c,d). The error bars for the 4.2~K data in panels a,b are within the symbols.}
\end{figure}

\begin{figure}[b]
\includegraphics[width=1\linewidth]{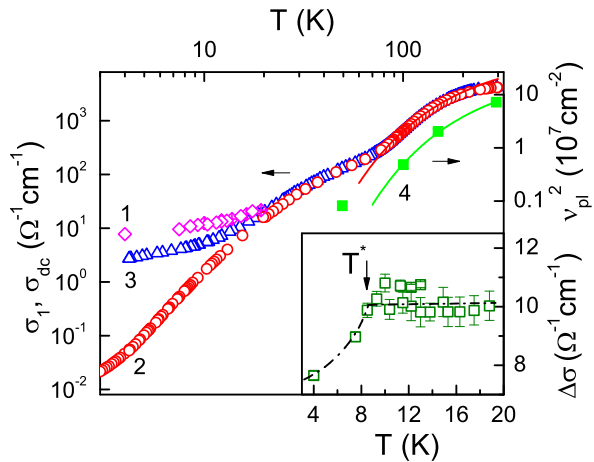}%
\caption{\label{Fig1} (Color online) Terahertz conductivity $\sigma_1$($T$) of FeSi measured at $\nu$=21 cm$^{-1}$ (1) in comparison with the static conductivity $\sigma_\mathrm{dc}$($T$) of etched (2) and polished (3) samples. Activation asymptotics of $\sigma_\mathrm{dc}$ and the squared plasma frequency $\nu_\mathrm{pl}^2$ (4) $\sigma_\mathrm{dc}$,$\nu_\mathrm{pl}^2$$\propto$exp(-$E_\mathrm{g}$/2$k_B$$T$)  are shown by solid lines. Inset presents the difference $\Delta$$\sigma$($T$)=$\sigma_1$($T$)--$\sigma_\mathrm{dc}$($T$). Arrow points to a kink on $\Delta$$\sigma$($T$) at $T^{*}$$\sim$8~K.}
\end{figure}

Figure~\ref{Fig2} presents the spectra of $\sigma_1(\nu)$ and  $\epsilon_1(\nu)$ directly calculated \cite{Gorshunov99} 
from the frequency dependent transmission $Tr(\nu)$  (panel c) and phase shift $\varphi(\nu)/\nu$ (panel d). Due to the large refractive index of $n=(\epsilon_1)^{1/2}\approx 14.5-15.8$ and relatively low absorption the interference within these very thin samples causes well resolved fringes. The dispersion of the ac-conductivity in the terahertz range from 17 to 40~cm$^{-1}$ [Fig.~\ref{Fig2}(b)] is small compared to that caused by the tails of optical conductivity expected from interband transitions or infrared optical phonons \cite{Cardona}. The optical conductivity decreases even further when $T$ falls below  20~K [Fig.~\ref{Fig2}(b)]. The comparison of $\sigma_1(T)$ with the dc data is plotted in Fig.~\ref{Fig1}. At low temperatures the terahertz conductivity saturates at a rather high value of approximately $10~(\Omega {\rm cm})^{-1}$; about two orders of magnitude above the $dc$ conductivity. 
Note the rather large difference between the $dc$ conductivity of etched and polished sample (curves 2 and 3 in Fig.~\ref{Fig1}). It demonstrates that surface defects can mask the intrinsic properties of FeSi in optical, transport and photoemission studies.

\begin{figure}[b]
\includegraphics[width=1\linewidth]{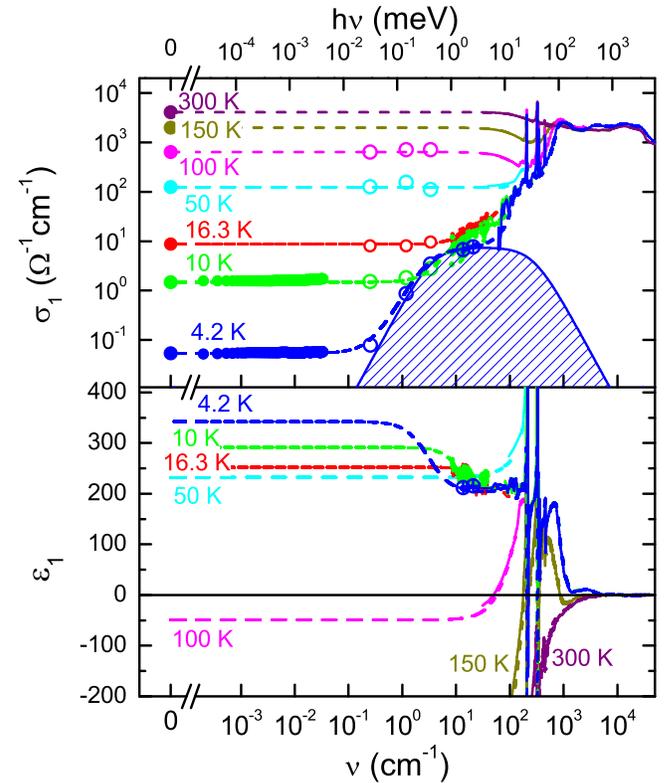}%
\caption{\label{Fig3} (Color online) Wide range optical conductivity $\sigma_1$ and dielectric function $\epsilon_1$ of FeSi at temperatures $4.2-300$~K. Dashed lines represent the results of least-square fitting by Eq.~(\ref{Eq1}). The $\sigma_\mathrm{dc}$ values are shown by filled circles at zero frequency in upper panel. Shaded area shows the terahertz resonance contribution at $T=4.2$~K with the parameters given in Tab.~\ref{table1}.}
\end{figure}

An essential information about low-energy electro\-dynamics of FeSi can be extracted from the wide-range spectra of $\sigma_1$($\nu$) and $\epsilon_1$($\nu$) plotted in Fig.~\ref{Fig3}. The evolution of $\sigma_1$($\nu$) shows that the Drude contribution becomes gradually weaker as the temperature is lowered; below 100~K no Drude roll-off can be resolved in the gap any more. Correspondingly the low-frequency dielectric constant becomes positive at $T<100$~K; this provides evidence that the compound enters an insulating state at low temperature. Most important for the present Communication, when $T$ drops below 20~K $\sigma_1$($\nu$) exhibits a wide step-like feature between 0.1 and 10~cm$^{-1}$ that is accompanied by a corresponding step in $\epsilon(\nu)$.

\begin{figure}
\includegraphics[width=1\linewidth]{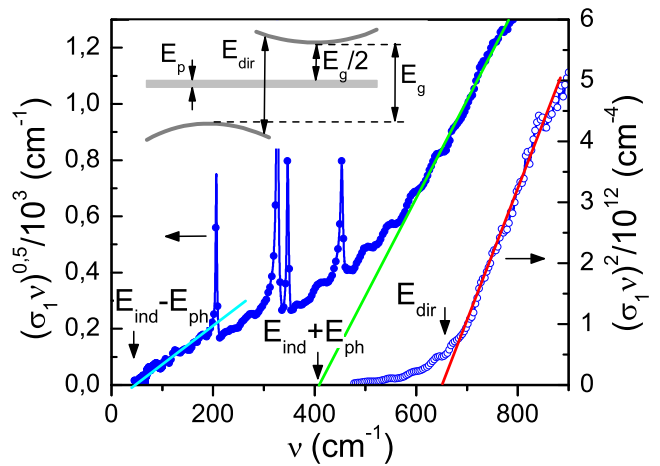}%
\caption{\label{Fig4} (Color online) The scaling of $\sigma_1(\nu, T=4.2~{\rm K})$ data by the relations $(\sigma_1 \nu)^{2}\propto(h\nu -E_\mathrm{dir})$ and $(\sigma_1\nu)^{0.5}\propto (h\nu \pm E_\mathrm{ph}- E_\mathrm{ind})$ allows for determination of the direct and indirect gaps (see text) \cite{Cardona}. Solid lines correspond to the least-squares fits. The proposed band structure of FeSi is sketched in the inset.}
\end{figure}

Applying the scaling relations \cite{Cardona}  $(\sigma_1\nu)^{2}\propto(h\nu-E_\mathrm{dir})$ and $(\sigma_1\nu)^{0.5}\propto(h\nu\pm E_\mathrm{ph}-E_\mathrm{ind})$ (here $E_\mathrm{ph}$ is the characteristic phonon energy), 
we obtain the direct and indirect energy gaps at $T$=4.2~K as depicted in Fig.\ref{Fig4}. The value of $E_\mathrm{dir}\approx 650~{\rm cm}^{-1}$ (corresponding to 81~meV) confirms previous reports \cite{Schles93,Menzel2009}. However, combining  $E_\mathrm{ind}-E_\mathrm{ph}\approx44~{\rm cm}^{-1}$ (5.5~meV) with $E_\mathrm{ind}+E_\mathrm{ph}\approx410~{\rm cm}^{-1}$ (51~meV) results in a rather low value of $E_\mathrm{ind}\approx227~{\rm cm}^{-1}$ (28.4~meV), which is only half of the energy gap $E_\mathrm{g}=60$~meV reported previously based on transport measurements \cite{Schles93,Sluchanko2000,Corti03} (Fig.\ref{Fig1}). Note also that the obtained characteristic phonon frequency $E_\mathrm{ph}=183~{\rm cm}^{-1}$ agrees very well the $E$ mode ($180~{\rm cm}^{-1}$) that shows the largest broadening with increasing temperature due to strong electron-phonon interaction \cite{Racu07}.

In order to clarify the discrepancy between transport gap ($E_\mathrm{g}\approx 60$~meV) and indirect optical gap ($E_\mathrm{ind}\sim 30$~meV) we fit the conductivity and dielectric data, $\sigma_1(\nu)$ and $\epsilon_1(\nu)$, by the sum of Drude and Lorentz terms \cite{Dressel}:
\begin{eqnarray}
\sigma(\nu)=\frac{\nu_\mathrm{pl}^2}{\gamma_{0}-i\nu}+\frac{1}{2}\sum_{i=1}^N \frac{\Delta\epsilon_\mathrm{i}\nu_\mathrm{0i}^2\nu}{\gamma_\mathrm{i}\nu+i(\nu_\mathrm{0i}^2-\nu^2)}
\label{Eq1}
\end{eqnarray}
where $\nu_\mathrm{pl}$, and $\gamma_0$ are plasma frequency and scattering rate of charge carriers, $\nu_\mathrm{0i}$, $\Delta$$\epsilon_\mathrm{i}$, and $\gamma_\mathrm{i}$ are eigenfrequency, dielectric contribution and damping (for $i$th Lorentzian), respectively. The fits are shown by dashed lines in Fig.~\ref{Fig3}. The number of Lorentz contributions $N$ was chosen minimal to describe the dispersion due to optical phonons and interband transitions; the estimated parameters of the features agree well with previous attempts \cite{Schles93,Menzel2009}. The Drude contribution to the fit yields a plasma frequency $\nu_\mathrm{pl}$ that follows the activation law $\nu_\mathrm{pl}^{2}\propto \exp\{-E_\mathrm{g}/2 k_B T\}$ with $E_\mathrm{g}\sim 70$~meV for $T>100$~K (Fig.~\ref{Fig1}). 
This observation proves that charge carriers are thermally excited from the lower to the upper band. The energy gap $E_\mathrm{g}$ of FeSi eventually becomes irrelevant but its size does not change with increasing temperature. 
Along these lines, the lower value of $E_\mathrm{ind}\sim 30$~meV corresponds to a finite density of states inside the gap rather than to a shift of bands. This suggestion is supported by the broad absorption identified in the terahertz conductivity for $T<20$~K (Fig.~\ref{Fig3}). Comparing the Lorentzian parameters listed in Tab.~\ref{table1} we see 
that  $\Delta\epsilon$ and $\gamma$ increase with lowering temperature while $\nu_0$ stays approximately constant ($\nu_0=4 \pm 0.5$~meV). Note that the enormous increase of the damping from $\gamma\approx 3.7$~meV to 35~meV when the temperature is reduced from 16 to 4.2~K rules out the assignment of this broad feature to excitations from extrinsic defects or impurity band.

In order to obtain a better insight into the  origin of the anomalous terahertz contribution to the optical response, let us consider the excess conductivity $\Delta\sigma(T)=\sigma_1(T)-\sigma_\mathrm{dc}(T)$ plotted in the inset of Fig.~\ref{Fig1}. 
At temperatures $8~{\rm K}<T<20$~K $\Delta\sigma(T)=10\pm1~(\Omega {\rm cm})^{-1}$ is approximately constant; but below $T^{*}\approx 8$~K it drops rapidly down to $\Delta\sigma(4.2~{\rm K})\approx7.5~(\Omega{\rm cm})^{-1}$. It is important to note that the value of $T^{*}$ is very close to the ``freezing'' temperature of interacting quasiparticles $T_\mathrm{m}\approx 7$~K found previously from transport and magnetic studies \cite{Glushkov2004}. These quasiparticles (i.e.\ spin polarons) correspond to the states of the many-body resonance (of width $E_\mathrm{p}\sim 6$~meV), which appears at the Fermi level below 80~K due to on-site Hubbard repulsion \cite{Sluchanko2000}. This now explains  the smaller value of the indirect gap $E_\mathrm{ind}\approx E_\mathrm{g}/2$ that defines the position of this resonance in the middle of the energy gap (see inset in Fig.~\ref{Fig4}).

\begin{table}[b]
\caption{\label{table1}%
The parameters of Lorentz terms used to fit the terahertz absorption feature in $\sigma(\nu$) of FeSi at different temperatures $T$. $\Delta\epsilon=(\nu_\mathrm{pl}/\nu_0)^2$ describes the oscillator strength, $\nu_0$ the center frequency, and $\gamma$ the damping parameter.}
\begin{ruledtabular}
\begin{tabular}{ccccc}
$T$~(K)&
$\Delta\epsilon$&
$\nu_0$ (cm$^{-1}$)&
$\gamma$ (cm$^{-1}$)\\
\colrule
4.2 & $107 \pm 30$ & $32 \pm 5$ & $280 \pm 30$\\
10 & $94 \pm 10$ & $36 \pm 10$ & $280 \pm 40$\\
16.3 & $28 \pm 5$ & $25 \pm 5$ & $30 \pm 10$\\
\end{tabular}
\end{ruledtabular}
\end{table}
We can conclude that in FeSi due to strong on-site Coulomb repulsion below $T\approx 80$~K a many-body resonance appears in the electronic density of states right in the middle of the energy gap \cite{Sluchanko2000}. 
The exchange field of $H_{e}\approx 34\pm 6$~T splits the band, and spin excitations between the resultant electronic levels cause the anomalous terahertz contribution to the optical conductivity of FeSi.
The inherent resonance width $E_\mathrm{g}\approx 6$~meV corresponds to the binding energy of spin polaronic states \cite{Sluchanko2000}
The extremely large damping $\gamma \approx 280~cm^{-1} (\approx35$~meV) 
is an estimate of the spin-fluctuation rate. 

This approach allows us to interpret the terahertz feature in the $\sigma_1(\nu)$ spectra: since the energy of the mode is extremely low ($h\nu_0\ll E_\mathrm{ind}$), we probe excitation within the intra-gap many-body resonance.
Using the oscillator strength $\Delta\epsilon\nu_0^2=(1.1\pm 0.4)\times 10^{5}~{\rm cm}^{-2}$ (Tab.~\ref{table1}) and effective mass $m^{*}\approx m_0$, we can estimate the particles concentration $n_0=\pi m^*\Delta\epsilon\nu_0^2/e^2=(1.2\pm 0.4)\times 10^{18}~{\rm cm}^{-3}$ that contribute to the terahertz absorption at $T=4.2$~K. 
The remarkable agreement of $n_0$ with the effective concentration of the many-body states $n_\mathrm{sp} \approx 10^{17} - 10^{18}~{\rm cm}^{-3}$ \cite{Sluchanko2000,Glushkov2004} strongly supports
a spin-polaronic ground state of FeSi. 
The terahertz absorption can then be associated with intra-gap excitations between the electronic levels split by the exchange field $H_{e}=h\nu_0/(2\mu_B)=34\pm 6$~T, which agrees well with the value of $H_{e} =35 \pm 10$~T previously estimated from magnetic studies \cite{Sluchanko2002}. The very large damping $\gamma\approx 35$~meV at low temperatures indicates strong spin fluctuations that cause transitions between the excited spin states and the ground state with a relaxation time $\tau=(2\pi c \gamma)^{-1}\approx 1.9\times 10^{-14}$~s. Both, the low concentration $n_0$ and the large damping $\gamma$ below 10~K make it difficult to detect this feature in photoemission experiments \cite{Klein08,Arita2008}.

Furthermore, our finding gives us now the possibility to distinguish also between different types of intra-gap excitations observed in the far infrared spectra of other strongly correlated electron systems 
identified as Kondo insulators \cite{Aeppli92,Riseborough00}. 
Similar features have been observed in the $\sigma(\nu)$ spectra of SmB$_{6}$ \cite{Gorshunov99,SluchPRB00} and YbB$_{12}$ \cite{Gorshunov06} 
and are ascribed to exciton-polaronic states 
that appear in the gap due to $4f - 5d$ charge fluctuations. 
The parameters of the Lorentz terms extracted for SmB$_{6}$ ($\nu_{0}\approx 24~{\rm cm}^{-1}$,  $\Delta\epsilon\approx 12$ and $\gamma \approx 7~{\rm cm}^{-1}$ \cite{Gorshunov99,SluchPRB00}) and YbB$_{12}$ ($\nu_{0}\approx 22~{\rm cm}^{-1}$,  $\Delta\epsilon \approx 75$ and $\gamma \approx 15~{\rm cm}^{-1}$ \cite{Gorshunov06}) characterize the exciton-polaronic complexes that arise from the coupling of itinerant electrons to soft valence fluctuations \cite{Kikoin95}. 
The characteristic times of charge fluctuations ($\tau \propto \gamma^{-1} \approx (4-8) \times 10^{-13}$~s for SmB$_{6}$ and YbB$_{12}$) and spin fluctuations ($\tau\propto 1.9\times 10^{-14}$~s for FeSi) are significantly different because of the different origin of the ground states. Further theoretical treatments are required to thoroughly understand this novel aspect of strongly correlated electron systems.

\begin{acknowledgments}
The authors are grateful to Prof. A.A.Menovsky for the grown FeSi single crystals. This work was supported by RAS Programme ``Strongly Correlated Electrons'' and Federal Programme ``Scientific and Educational Human Resources of Innovative Russia''. \end{acknowledgments}

\bibliography{FeSi2}

\end{document}